\documentclass[11pt]{article}
\usepackage{moriond,epsfig}

\def\Journal#1#2#3#4{{#1} {\bf #2}, #3 (#4)}

\def\NIMA{{\em Nucl. Instrum. Methods} A}

\def\be{\begin{equation}}
\def\ee{\end{equation}}
\def\bea{\begin{eqnarray}}
\def\eea{\end{eqnarray}}

\begin{document}
\vspace*{4cm}
\title{Coincidence analysis in ANTARES: Potassium-40 and muons}

\author{D. Zaborov on behalf of the Antares collaboration}

\address{Institute for Theoretical and Experimental Physics\\
Moscow, 117218, Russia 
}

\maketitle\abstracts{
A new calibration technique using natural background light of sea water
has been recently developed for the ANTARES experiment.
The method relies on correlated coincidences produced in triplets of optical modules 
by Cherenkov light of $\beta$ particles originated from $^{40}$K decays.
A simple but powerful approach to atmospheric muon flux studies is currently being developed
based on similar ideas of coincidence analysis.
This article presents the two methods in certain detail and explains their role in the
ANTARES experiment.
}

\section{Introduction}
ANTARES is a large water Cherenkov detector operating in Mediterranean sea
40 km offshore Toulon (France) at the depth of 2470 m.\cite{Proposal,Status}
An ANTARES storey includes a triplet of optical modules\cite{OM} oriented at 45$^o$ downwards and outwards of the vertical axis.
Twenty five storeys, chained together with a step of 14.5 m, form a detector line.
Several calibration systems and techniques ensure a sub-nanosecond precision 
of the Cherenkov pulse measurements, which is required to achieve the high angular resolution of the neutrino telescope.
Important roles are played by in situ measurements using LED Beacons\cite{OpticalBeacons} and Potassium-40.

Potassium-40 is a $\beta$-radioactive isotope naturally present in sea water.
The energy freed in $^{40}$K decays amounts to 1.3 MeV,
that well exceeds Cherenkov threshold for electrons in water (0.26~MeV) and is sufficient to produce up to 150 Cherenkov photons.
If the decay occurs in the vicinity of a detector storey,
a coincident signal may be seen by two of the three optical modules constituting the triplet (local coincidence).
This effect and its use in the ANTARES experiment are explained in section \ref{sect:k40}.
In the case of two modules located on different storeys
the probability of a genuine coincidence from $^{40}$K is negligibly small.
Instead, a signal originated from atmospheric muons could become dominating, if was not overwhelmed by
the random background from $^{40}$K and bioluminescence.
In order to reduce the random background we require a local coincidence (within 20 ns time window)
at each of the two storeys rather than just a hit.
Section \ref{sect:muons} presents such an analysis for the case of adjacent storeys.

For the purpose of detailed $^{40}$K measurements a dedicated type of calibration runs is defined in ANTARES (K40 runs).
During a K40 run (typically 20 minutes) all local coincidences detected in any OM triplet of ANTARES are selected 
by a dedicated data filter algorithm, so-called K40 trigger, and saved on disk for later processing.
Importantly, this type of runs is also perfectly suitable for the adjacent floor coincidence studies.

\section{Calibration with Potassium-40}\label{sect:k40}
\begin{figure}
\epsfig{figure=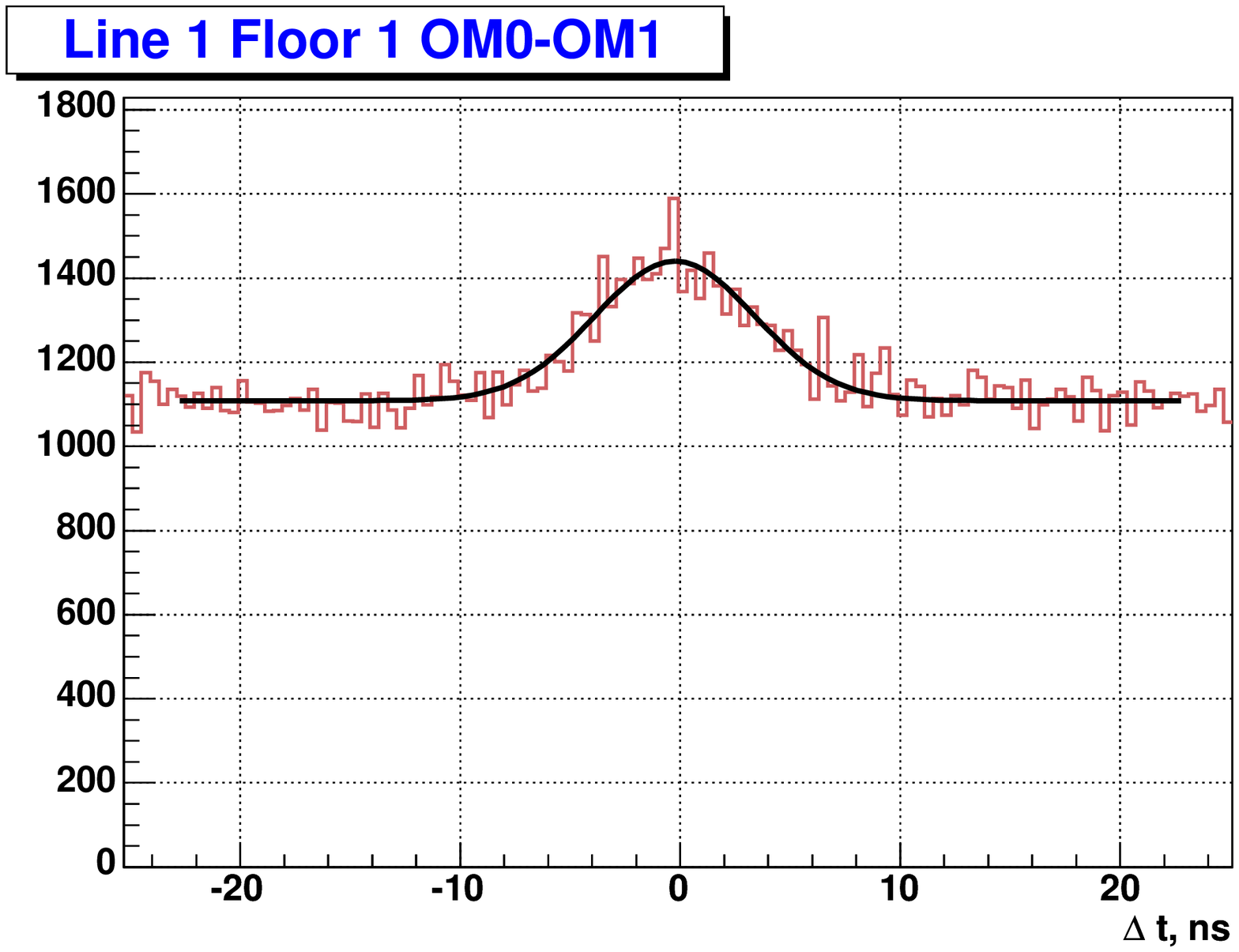,height=2.1in}
\epsfig{figure=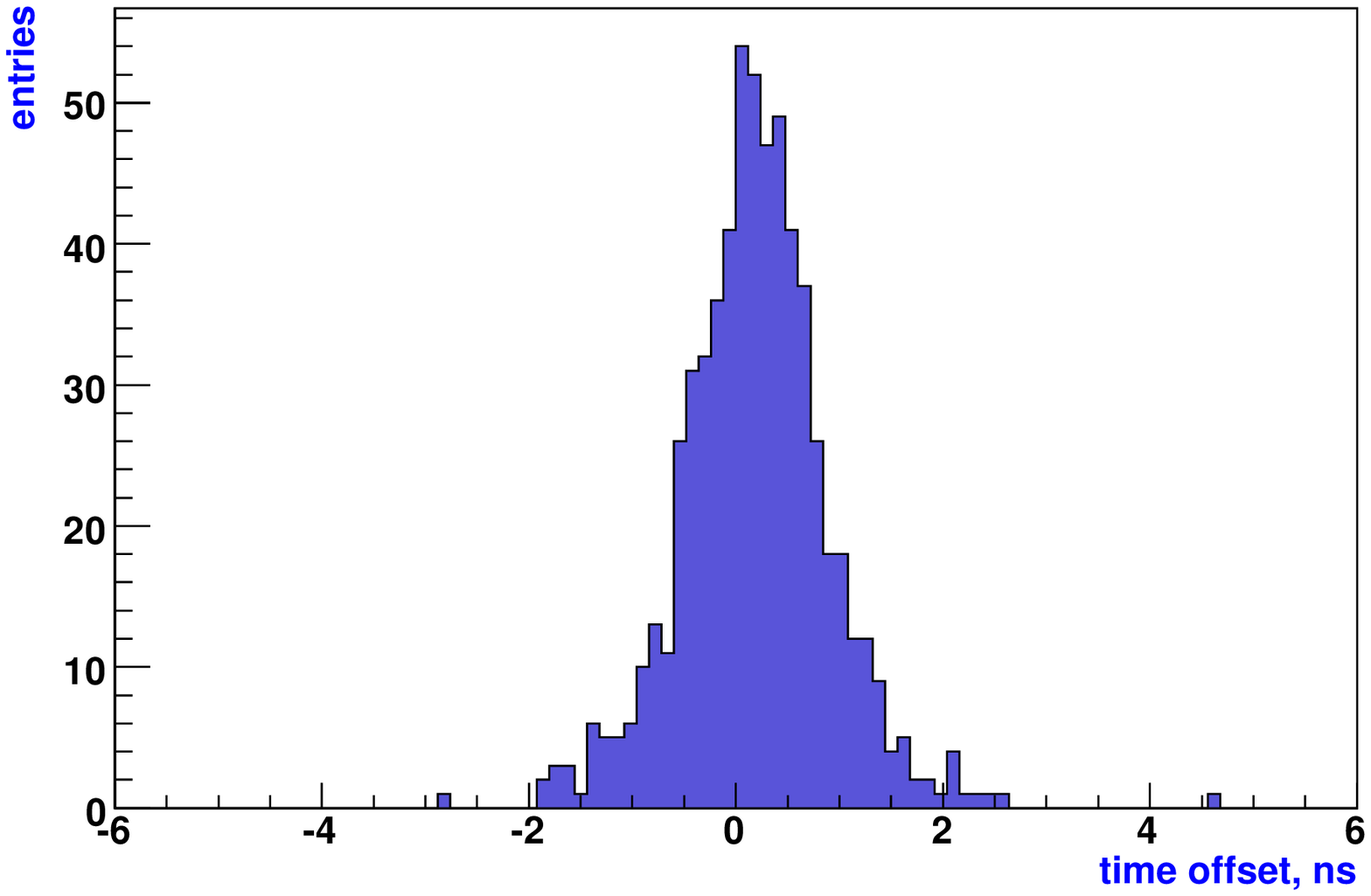,height=2.1in}
\caption{Left: time delay between hits in two optical modules of the same detector storey (example).
Coincidence peak from Potassium-40 is evident (see text).
Right: time offset of the coincidence peak (Line 1-10 data).}
\label{k40_peak}
\end{figure}

Due to the difference in positions of the optical modules (the distance between OM centers is 1.0 m)
the signals are detected by the two OMs with a noticeable delay.
When averaging over many events (locations of decayed nucleus) 
this results in a time spread of 4.0 ns (RMS),
that is experimentally observed (see Fig.~\ref{k40_peak}).
This value is in excellent accordance with Monte Carlo simulations.
As one can see from Fig.~\ref{k40_peak}, the coincidence peak is comparable in amplitude with the pedestal of random coincidences,
which are composed of truly uncorrelated background photons (from $^{40}$K or bioluminescent emission).
Ideally, if the optical modules are all identical, perfectly calibrated, and arranged in triplets symmetrically,
the coincidence peak must be aligned with zero.
Experimental measurements (see Fig.~\ref{k40_peak})
show a small spread of the offsets (0.68 ns RMS), that confirms the high accuracy of timing calibration in ANTARES
and gives a measure of disagreement between the $^{40}$K data and currently used calibration set.
The present results refer to so-called Dark Room calibration, which is performed on shore before line deployment.
An improvement is foreseen with the in-situ LED Beacon calibration, which is currently in progress.

\begin{figure}
\epsfig{figure=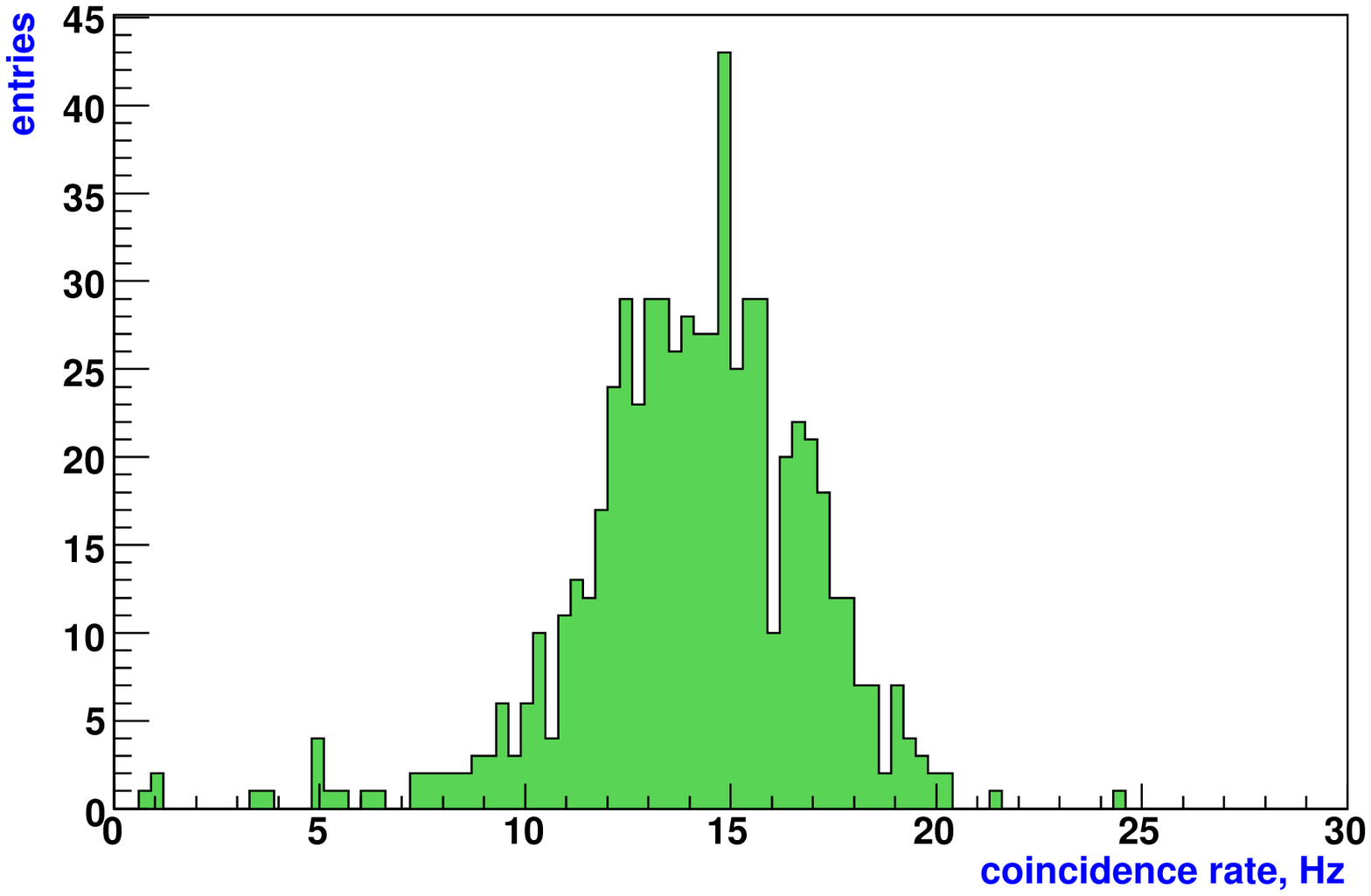,height=2.0in}
\epsfig{figure=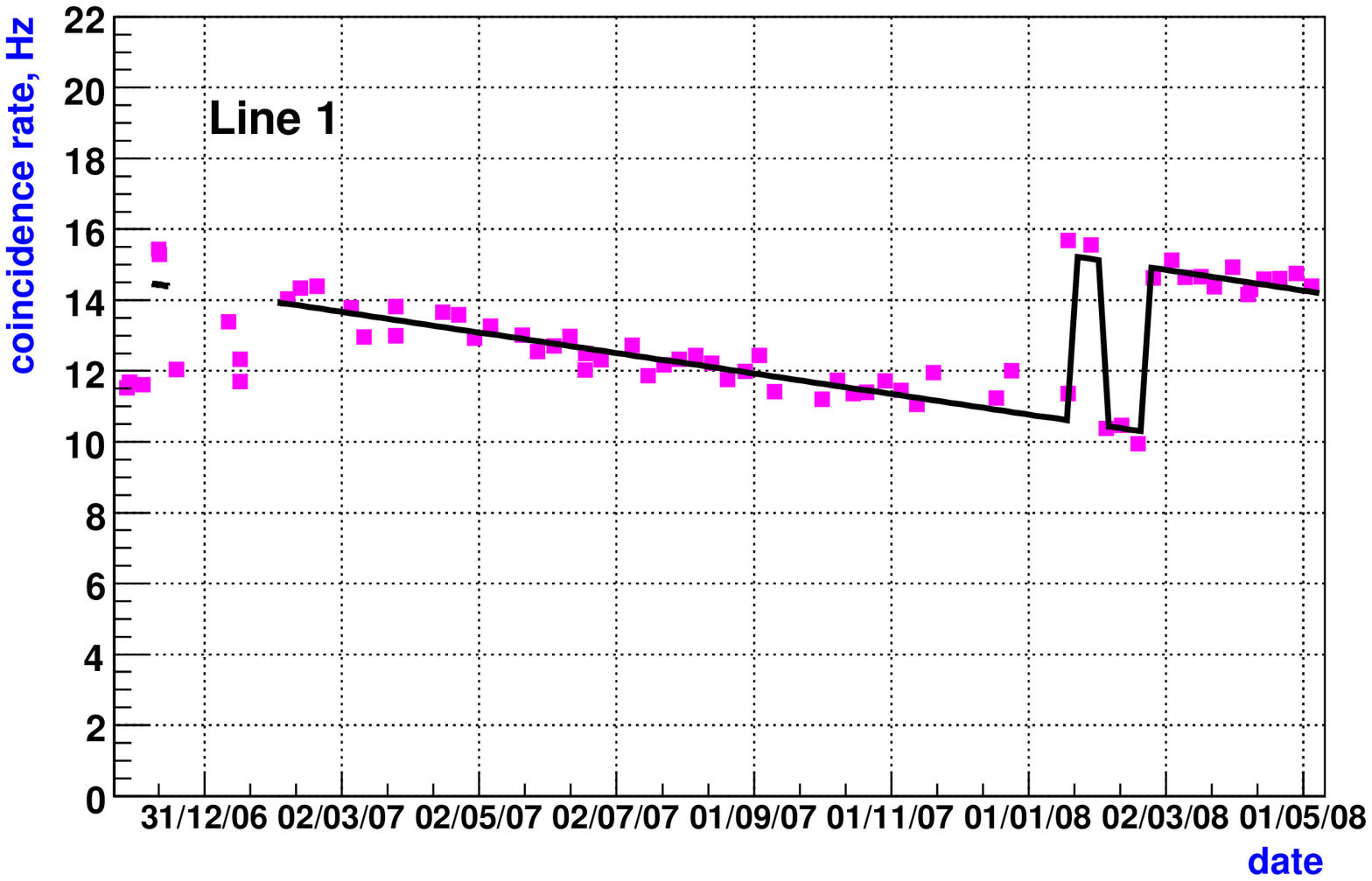,height=2.0in}
\caption{Left: rate of correlated coincidences produced in ANTARES storeys by $^{40}$K decays in sea water (Line 1-10 data).
Right: evolution of average rate of correlated coincidences with time (Line 1 data).
Continuous degradation of OM sensitivity is noticeable, as well as the effects of changing hardware thresholds.}
\label{rate_all}
\end{figure}

The rate of correlated coincidences can be defined as the integral under the coincidence peak (excluding pedestal)
normalized to the effective duration of observation period, and properly corrected for dead time
of the electronics and data acquisition.
We rely on a Gaussian fit to compute the rate.
The observed average value amounts to 14 Hz (see Fig. \ref{rate_all}, left),
that is in good agreement with MC simulations.
It has been checked experimentally that the coincidence rate is not affected by variations of 
the bioluminescent background.
This important observation suggests that the $^{40}$K measurements
can be used as a robust calibration tool.
In addition, it also provides a confirmation of single-photon character 
of the bioluminescent emission.
Indeed, if a bioluminescent process produced bunches of photons, correlated in nanosecond scale,
an increase of the correlated coincidence rate would be observed for high-background runs.

It can be shown by Monte Carlo simulation that the rate of correlated coincidences is proportional
to the detection efficiencies of both the optical modules involved.
Thus, for a triplet one can write three equations as follows:
\addtolength{\abovedisplayskip}{-3mm}
\addtolength{\belowdisplayskip}{-3mm}
\begin{equation}
\begin{array}{rcl}
 r_{12} & = & k s_1 s_2\\
 r_{23} & = & k s_2 s_3\\
 r_{31} & = & k s_3 s_1
\end{array}
\label{eq1}
\end{equation}
where $s_{1(2,3)}$ is the sensitivity of OM1(2,3) in arbitrary units,
$r_{12(23,31)}$ is the rate of correlated coincidences between OM1 and OM2 (OM2 and OM3, OM3 and OM1),
and $k$ is a normalization factor. 
We will assume $k$ constant, i.e. not varying from module to module or with time.
For a triplet of optical modules one can unambiguously solve system
(\ref{eq1}) and extract the three sensitivity coefficients $s_{1(2,3)}$.
Note that the absolute normalization of the OM sensitivities requires precise knowledge of the OM angular acceptance
(alternatively, OM angular acceptance can be constrained by the $^{40}$K measurements).
\addtolength{\abovedisplayskip}{3mm}
\addtolength{\belowdisplayskip}{3mm}

So far, OM sensitivity measurements have been done with $^{40}$K for ten detector lines of ANTARES,
helping to locate mistuned and degraded modules.
A significant drop of sensitivity was observed for some optical modules during first months of operation.
In average a decrease of coincidence rates by 2.5\% per month has been observed, 
that corresponds to about 1.2\% decrease in OM sensitivity per month (see Fig.~\ref{rate_all}, right).
These measurements agree qualitatively with observations of counting rates, which are continuously monitored in ANTARES
(but can be affected by bioluminescence).
The present method also allowed to measure the effect of tuning of hardware thresholds,
recently performed in ANTARES in order to compensate the sensitivity drop.
It was found that the adjustment of pulse-height discrimination thresholds allowed to
effectively recover the initial detector sensitivity (see Fig.~\ref{rate_all}, right).
This supports the hypothesis that the sensitivity drop occurs due to a continuous decrease
of gain of the photomultiplier tubes.
Further measurements with $^{40}$K are being performed to better investigate the observed phenomena
and maintain high accuracy of the OM sensitivity calibration.

\section{Low energy atmospheric muons}\label{sect:muons}
\begin{figure}
\epsfig{figure=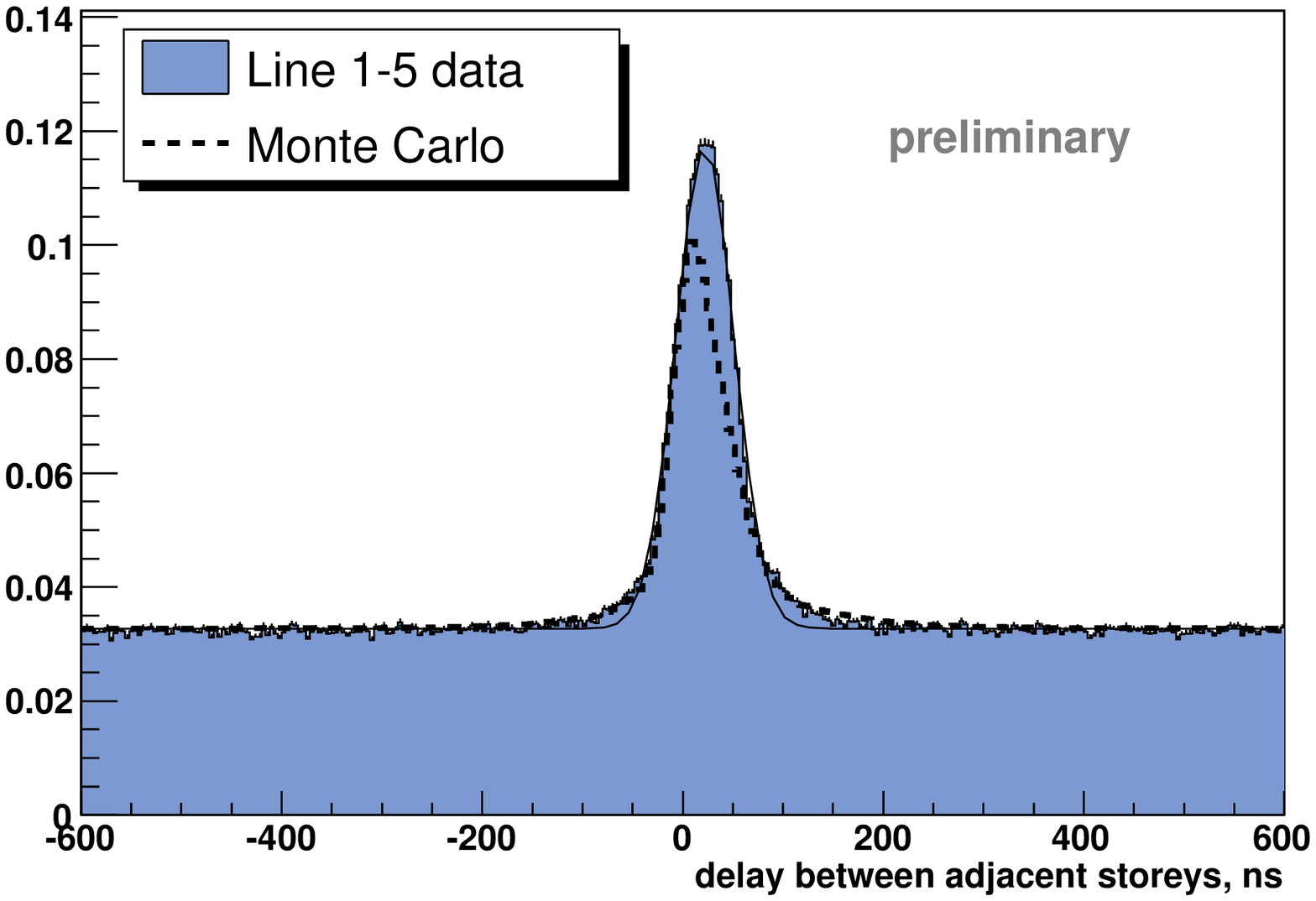,height=2.0in}
\epsfig{figure=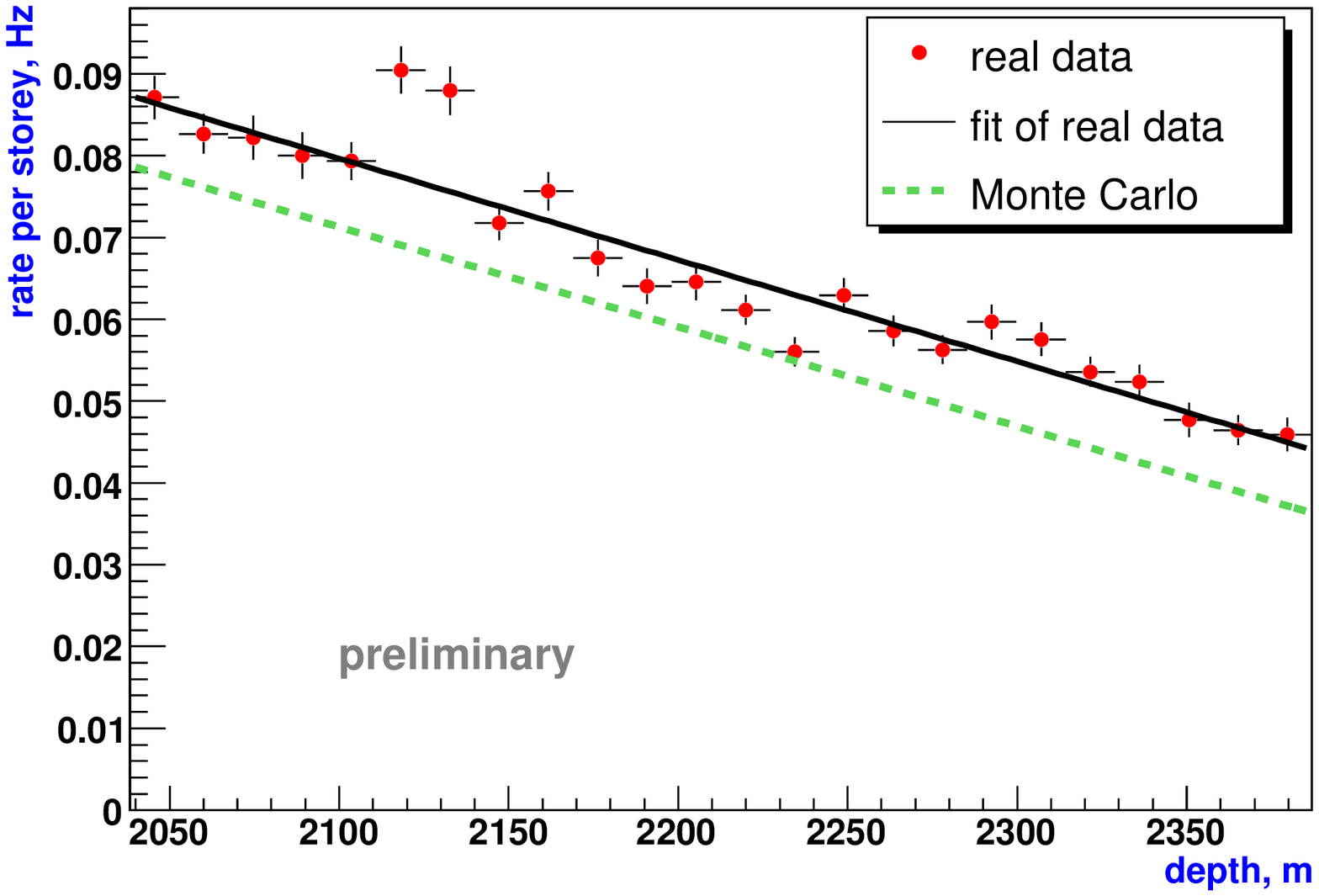,height=2.0in}
\caption{Left: time delay between two local coincidences detected in adjacent detector storeys
(A2 coincidence plot, example). Right: depth dependence of A2 coincidence rate (Line 1-5 data).
Each of the 24 points corresponds to a pair of adjacent detector floors.
Both plots are preliminary.
}
\label{delay_between_floors}
\end{figure}

An experimental plot of adjacent floor coincidences (A2 coincidences)
exhibits a prominent peak shifted toward positive values (see Fig.~\ref{delay_between_floors}).
This is the most basic signal of (downward-going) atmospheric muons detected so far in ANTARES.
A Monte Carlo simulation performed using MUPAGE\cite{MUPAGE} agrees with the observations.
The rate of correlated A2 coincidences (A2 rate), which can be defined 
as the integral under the peak, is directly linked to the atmospheric muon flux.
One should note however that a good knowledge of absolute efficiency of the optical modules,
as well as their angular response, is needed to accurately convert the A2 rate into the flux.
Interestingly, some important information on the angular response curve can be extracted
from the shape of the A2 plot.
A detailed study trying to exploit this possibility is currently ongoing.
Importantly, the analysis can be repeated for each detector floor separately, thus providing access to the depth-intensity relation of the muon flux.
A preliminary analysis (see Fig. \ref{delay_between_floors}, right) has demonstrated the effect of muon absorption with depth.
A dedicated effort is currently ongoing to develop the necessary corrections for the different sensitivities of the optical modules
(and therefore storeys), and further develop this technique.

Clearly this analysis does not allow to determine the properties of each event (e.g. muon zenith angle).
However, certain collective properties of the muon flux might well be accessible.
Moreover, any systematic errors and inefficiencies of likelihood maximization and event selection procedures are also avoided in this case.
In addition, the energy threshold of the coincidence study is much lower, thanks to the use of highly-efficient K40 trigger and 
very low requirements imposed on the number of signal hits by the analysis itself. 
Thus the coincidence study can not only provide a redundancy to muon flux measurements but also 
deliver a lot of useful and complementary information concerning the detector response and atmospheric muon flux.

\section{Outlook}
A new calibration technique grew up from an academic study of local coincidences in an ANTARES storey.
The method represents a unique tool for measurements of optical module sensitivities
in situ, and a valuable tool for time calibration in ANTARES.
Thanks to this method the effect of OM efficiency drop was precisely measured,
and a preliminary time calibration of ANTARES was confirmed in an independent way.
Presently the method is used on a regular basis to control the sensitivities of the optical modules.
A new approach to atmospheric muon flux studies with ANTARES was suggested,
which also relies on a simple coincidence analysis.
The method allows to measure the muon flux and, more importantly, its depth dependence.
Some preliminary results were presented in this paper, further analysis is ongoing.

\end{document}